\documentclass[letterpaper,twocolumn,prl,aps,superscriptaddress,amsmath,amssymb,floatfix]{revtex4-2}
\usepackage{mathptmx}
\usepackage[latin9]{inputenc}
\usepackage{color}
\usepackage{verbatim}
\usepackage{amssymb,amsthm,amsmath,amssymb,bbm}
\usepackage{adjustbox}
\usepackage{esint}
\usepackage[percent]{overpic}
\usepackage{graphicx,float,subfigure}
\usepackage{dcolumn}
\usepackage{bm,braket,color}

\usepackage[unicode=true,
bookmarks=true,bookmarksnumbered=false,bookmarksopen=false,
breaklinks=false,pdfborder={0 0 1},backref=false,colorlinks=true]
{hyperref}
\hypersetup{
	linkcolor=magenta,urlcolor=blue,citecolor=blue,pdfstartview={FitH},hyperfootnotes=false}
\makeatletter

\usepackage{textcomp}
\usepackage{epstopdf}

\pdfpageheight\paperheight
\pdfpagewidth\paperwidth

\@ifundefined{textcolor}{}{%
	\definecolor{BLACK}{gray}{0}
	\definecolor{WHITE}{gray}{1}
	\definecolor{RED}{rgb}{1,0,0}
	\definecolor{GREEN}{rgb}{0,1,0}
	\definecolor{BLUE}{rgb}{0,0,1}
	\definecolor{CYAN}{cmyk}{1,0,0,0}
	\definecolor{MAGENTA}{cmyk}{0,1,0,0}
	\definecolor{YELLOW}{cmyk}{0,0,1,0}
}

\usepackage{xcolor}
\usepackage{soul}
\definecolor{blue}{rgb}{0,0,1}
\definecolor{red}{rgb}{1,0,0}
\definecolor{green}{rgb}{0,1,0}

\DeclareMathAlphabet{\mathcal}{OMS}{cmsy}{m}{n}
\DeclareSymbolFont{largesymbols}{OMX}{cmex}{m}{n}

\usepackage{soul}

\makeatother
\begin{document}
\title{Kardar-Parisi-Zhang dynamics in an open integrable system: beyond the spontaneous-symmetry-breaking ansatz}
\author{Guo-Qiang Wang}
\affiliation{Laboratory of Quantum Information, University of Science and Technology
of China, Hefei, Anhui 230026, P. R. China}
\affiliation{Anhui Province Key Laboratory of Quantum Network, University of Science and Technology of China, Hefei 230026, P. R. China}
	
\author{Chang-Ling Zou}
\affiliation{Laboratory of Quantum Information, University of Science and Technology
of China, Hefei, Anhui 230026, P. R. China}
\affiliation{Anhui Province Key Laboratory of Quantum Network, University of Science and Technology of China, Hefei 230026, P. R. China}
\affiliation{CAS Center for Excellence in Quantum Information and Quantum Physics, University of Science and Technology of China,
Hefei 230026, China}
\affiliation{Hefei National Laboratory,
University of Science and Technology of China, Hefei 230088, China}

	\author{Guang-Can Guo}
\affiliation{Laboratory of Quantum Information, University of Science and Technology
of China, Hefei, Anhui 230026, P. R. China}
\affiliation{Anhui Province Key Laboratory of Quantum Network, University of Science and Technology of China, Hefei 230026, P. R. China}
\affiliation{CAS Center for Excellence in Quantum Information and Quantum Physics, University of Science and Technology of China,
Hefei 230026, China}
\affiliation{Hefei National Laboratory,
University of Science and Technology of China, Hefei 230088, China}

\author{Xu-Bo Zou}
\email{xbz@ustc.edu.cn}
\affiliation{Laboratory of Quantum Information, University of Science and Technology
of China, Hefei, Anhui 230026, P. R. China}
\affiliation{Anhui Province Key Laboratory of Quantum Network, University of Science and Technology of China, Hefei 230026, P. R. China}
\affiliation{CAS Center for Excellence in Quantum Information and Quantum Physics, University of Science and Technology of China,
Hefei 230026, China}
\affiliation{Hefei National Laboratory,
University of Science and Technology of China, Hefei 230088, China}

	\begin{abstract}
    The universality of dynamical scaling laws constitutes a cornerstone in the theoretical understanding of quantum many-body systems, particularly in non-equilibrium settings. Recent advancements have proposed a phenomenological ansatz based on spontaneous symmetry breaking (SSB) to unify the description of charge transport in open quantum systems. However, it remains unclear under which conditions it fails to capture the emergent hydrodynamics and if it does break down, whether nontrivial dynamics emerge. In this work we show that Kardar-Parisi-Zhang (KPZ) dynamics in an open integrable model (the B3 model), rather than diffusion from SSB, emerges. We find that the B3 model is equivalent to two interacting asymmetric XXZ spin chains and the ansatz can only capture the influence of the inter-chain interactions. When the initial state is appropriate, the asymmetric XXZ structure dominates the dynamics, which gives KPZ scaling behavior even when the hopping rate becomes negative. Our work motivates theory of charge transport in open systems beyond the ansatz based on SSB.
	\end{abstract}
	\maketitle
	
\smallskip{}
\noindent	\textit{\textbf{Introduction}}.---
	At the macroscopic scale, many-body quantum systems with conserved charge often exhibit universality--- identical hydrodynamic behavior may emerge from microscopically distinct models. This universality motivates us to understand quantum many-body physics from a general perspective rather than relying on specific microscopic models. In isolated systems, significant progress has been made, including effective field theory~\cite{EFT_1,EFT_2}, diffusive charge transport in chaotic systems~\cite{Chaotic_1}, subdiffusive dynamics in fracton systems~\cite{Fracton_1,Fracton_2,Fracton_3,Fracton_4} and superuniversality of superdiffusion in integrable models~\cite{Integrable_1,Integrable_2,Integrable_3,Integrable_4,Integrable_5}. Recently, growing attention has been focused on the universality in open systems, with substantial progress achieved both in analytical~\cite{Doyon_1,KPZ_open_1,Jin_1,Jin_2,Rabl,Ren_1,BKA_1} and phenomenological methods~\cite{Open_unif,SSB_the}. Among these advancements, the spontaneous symmetry breaking ansatz (SSBA)~\cite{Open_unif} is particularly important, as it provides a simple yet powerful framework to unify charge transport in systems governed by the Lindblad master equation~\cite{Lind}.
	
	At its core, SSBA utilizes the Feynman-Bijl variational wave functions~\cite{FB_1,FB_2,FB_3} to approximate low-energy excitations in the vicinity of steady states. Since long-time charge dynamics in open quantum systems is governed by steady states and their low-energy excitations, SSBA directly gives transport properties without requiring exact solutions to the model. This approach predicts the existence of universal diffusive dynamics in open systems with only local interactions and no constraints. Moreover, it has successfully captured the transport behaviors of long-range interacting models~\cite{Long_1,Long_2} and even those of constrained systems~\cite{Fracton_1,Fracton_2}. However, SSBA neglects the effects of terms in the master equation that drive variational wave functions out of the variational subspace. It remains unclear whether this approximation is valid, and if not, whether anomalous dynamics emerge.
	
	In this Letter, we report the failure of SSBA in the B3 model~\cite{YB_1,YB_2,YB_3}, where charge dynamics exhibit Kardar-Parisi-Zhang (KPZ)~\cite{KPZ_ori} scaling behavior instead of the diffusive dynamics predicted by SSBA. Specifically, the B3 model can be transformed into two interacting spin chains while SSBA only captures the influence of interchain interactions. However, when the system is prepared in an appropriate initial state, the interchain interaction effectively becomes irrelevant to charge dynamics. Consequently, each chain is equivalent to the asymmetric XXZ model~\cite{AS_XXZ_1,AS_XXZ_2,KPZ_sta} and KPZ dynamics emerge. It is worth noting that although the B3 model is Yang-Baxter integrable, the origin of its KPZ dynamics differs from that in isolated integrable systems such as the quantum Heisenberg magnet~\cite{Integrable_1,Integrable_2,Integrable_3}. Our work demonstrates that for a comprehensive theory of emergent hydrodynamics in open systems, a simplistic understanding based on SSB alone is insufficient.
	
	\begin{figure*}[t]
		\begin{center}
			\subfigure{
				\includegraphics{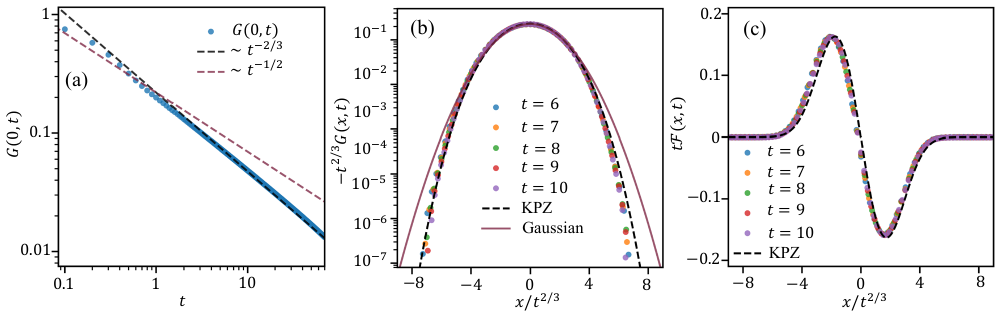}
			}
			\caption{
				\label{fig_B3_lind} \textbf{Charge transport of B3 model.}  (a) $G(x=0,t)$ for $t<70$. The black dashed line is $\sim t^{-2/3}$ and the red dashed line is $\sim t^{-1/2}$. (b) Rescaled $G(x,t=6,7,8,9,10)$. The black dashed line is from KPZ scaling function: $ bf_{\mathrm{KPZ}}(ax/t^{2/3})$ where $a=b\simeq0.426$. The red solid curve is $f_G(\xi)=\frac{1}{\sqrt{2\pi D}}\mathrm{exp}\left(-\frac{\xi^2}{2D} \right)$ where $ D\simeq3.125$, which is the closest Gaussian distribution to the scaling function of ${G}(x,t)$. Moreover, when $|x/t^{2/3}|\gtrsim8$, $-G(x,t)$ becomes negative and are not shown here. (c) Rescaled $\mathcal{F}(x,t=6,7,8,9,10)$. The black dashed line is from KPZ scaling function: $ \frac{2bx}{3at^{2/3}}f_{\mathrm{KPZ}}(ax/t^{2/3})$. Parameters: chain length $N=256$, $\gamma=1.2$ and bond dimension $\chi=64$.
			}
		\end{center}
	\end{figure*}
	
\smallskip{}
\noindent	\textit{\textbf{Breakdown of SSBA}}.---
	We consider a spin-1/2 chain of length $N$ under periodic boundary conditions, where the dynamics is governed by the Lindblad master equation
	\begin{eqnarray}
		\frac{\mathrm{d}\rho}{\mathrm{d}t}=\sum_{j=1}^{N}\left(-i[h_{j,j+1},\rho]+\mathcal{D}_{j,j+1}[\rho]\right),
		\label{master eqation}
	\end{eqnarray}
	with Hamiltonian density $h_{j,j+1}$ and dissipator $\mathcal{D}_{j,j+1}[\rho]=-\frac{1}{2}\{l_{j,j+1}^\dagger l_{j,j+1},\rho\}+l_{j,j+1}\rho l_{j,j+1}^\dagger$ ($l_{j,j+1}$ is the jump operator) acting on sites $j$ and $j+1$. We assume the system possesses a conserved $U(1)$ charge $Q=\sum_j q_j$ satisfying $[Q,h_{j,j+1}]=[Q,l_{j,j+1}]=0$, which implies a continuity relation $\mathrm{d}q_j(t)/\mathrm{d}t+J_{j}(t)-J_{j-1}(t)=0$ in Heisenberg picture, with $q_j$ and $J_j$ denoting the charge and current densities. Our goal is to understand the dynamical universality class governing charge transport, characterized by the expected value for charge density $C_{\rho,j}(t)=\mathrm{Tr}(q_j\rho(t))$ and current density $F_{\rho,j}(t)=\mathrm{Tr}(J_j\rho(t))$.
	
	For convenience, we apply the Choi isomorphism~\cite{Choi} to Eq.~(\ref{master eqation}), mapping operators ($\mathcal{O}=\sum_{a,a'}\mathcal{O}_{a,a'}\ket{a}\bra{a'}$) to vectors in a doubled Hilbert space as $|\mathcal{O})=\sum_{a,a'}\mathcal{O}_{a,a'}\ket{a}_l\ket{a'}_r$, where $\{\ket{a}\}$ is a basis of the system's Hilbert space. Under this mapping, the system can be intuitively interpreted as two chains of spins, and Eq.~(\ref{master eqation}) becomes ${\mathrm{d}|\rho)}/{\mathrm{d}t}=\mathcal{L}|\rho)$, where the Liouvillian superoperator $\mathcal{L}=\sum_j\mathcal{L}_{j,j+1}$ decomposes into three distinct contributions:
	\begin{equation}
		\mathcal{L}_{j,j+1}={A_{j,j+1}\otimes\mathbf{1}}+{\mathbf{1}\otimes B_{j,j+1}}+{l_{j,j+1}\otimes l_{j,j+1}^*},
		\label{Model_doub}
	\end{equation}
	with
	\begin{eqnarray}
		&& A_{j,j+1}=-ih_{j,j+1}-\frac{1}{2}l_{j,j+1}^\dagger l_{j,j+1},\\
		&&B_{j,j+1}=ih_{j,j+1}^T-\frac{1}{2}l_{j,j+1}^T l_{j,j+1}^*.
	\end{eqnarray}
	This decomposition reveals the structure of open-system dynamics in doubled space. The first two terms, $A\otimes\mathbf{1}$ and $\mathbf{1}\otimes B$, generate independent non-Hermitian evolution on the left and right chains respectively. The third term $l\otimes l^*$ couples the two chains and represents quantum jumps that create correlations between them. Physically, while the single-chain terms describe how the density matrix's row and column indices evolve separately, the inter-chain coupling captures the genuinely dissipative processes that mix different matrix elements. This separation will prove crucial: we shall demonstrate that under special conditions, the inter-chain coupling can become dynamically irrelevant.
	
    SSBA constructs the gapless Nambu-Goldstone modes in Heisenberg picture using Feynman-Bijl variational low-energy states
	\begin{equation}
		|m_k)=\frac{1}{2}\sum_{j=1}^N e^{ikj}(q_{j,l}+q_{j,r})|P_m),
	\end{equation}
	which are interpreted as spin waves with momentum $k$ and $P_m$ is the projection to $Q_l=Q_r=m$~\cite{Open_unif}. Within this variational subspace \{$|m_k)$\}, the charge dynamics takes the form
	\begin{equation}
		C_{\rho,j}(t)\simeq \frac{1}{N}\sum_{m=0}^{N}\sum_k e^{-ikj-E_k t}(\rho_0|m_k),
		\label{C_SSB}
	\end{equation}
	where $E_k$ is the eigenvalue of $\mathcal{L}^\dagger$ restricted to $|m_k)$. For generic local Lindbladians, $E_k\propto k^2$, yielding diffusive transport with dynamical exponent $z=2$.
	
	However, a close examination of the variational eigenvalue indicates the break down of SSBA. Taking $q_j=\sigma_j^z$ ($\sigma^{x,y,z}_j$ are Pauli operators and $\sigma_j^\pm=(\sigma_j^x\mp i\sigma_j^y)/2$), we find that for any $k$ and $k'$
	\begin{eqnarray}
		(m_{k'}|A_{j,j+1}^\dagger\otimes\textbf{1}|m_k)=(m_{k'}|\textbf{1}\otimes B^\dagger_{j,j+1}|m_k)=0,
	\end{eqnarray}
	which means that $A_{j,j+1}$ and $B_{j,j+1}$ can drive $|m_k)$ out of the variational state subspace $\{|m_k)\}$ without contributing to $E_k$. Consequently, the $k^2$ dispersion and the predicted $z=2$ diffusion arise from the interchain coupling term, while the effects generated by single-chain terms ($A_{j,j+1}$ or $B_{j,j+1}$) are excluded by the SSBA. When the charge dynamics generated by the single chain terms gives distinct diffusive behavior from $z=2$, then SSBA may fail to capture the true dynamical universality class.

	\begin{figure*}[t]
		\begin{center}
			\subfigure{
				\includegraphics{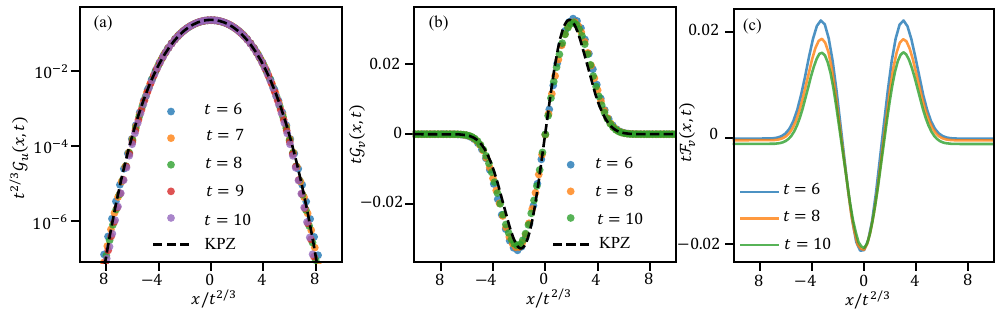}
			}
			\caption{
				\label{fig_2} \textbf{Origin of KPZ dynamics.} (a) Rescaled $\mathcal{G}_u(x,t=6,7,8,9,10)$. The black dashed line is from KPZ scaling function: $ bf_{\mathrm{KPZ}}(ax/t^{2/3})$ where $a=b\simeq0.426$. (b) Rescaled $\mathcal{G}_v(x,t=6,8,10)$. The black dashed line is $-\frac{2cx}{3dt^{2/3}}f_{\mathrm{KPZ}}(dx/t^{2/3})$ with $c\simeq0.08,\ d\simeq0.4$. (c) Rescaled $\mathcal{F}_v(x,t=6,8,10)$.
			}
		\end{center}
	\end{figure*}

\smallskip{}
\noindent \textit{\textbf{B3 Model}}.---
To demonstrate the failure of SSBA, we adapt the Yang-Baxter integrable open quantum system (B3 model)~\cite{YB_2,YB_3}, and show that the inter-chain coupling becomes dynamically irrelevant while the single-chain terms dominate the long-time behavior for appropriate initial states. The B3 model Hamiltonian density reads
	\begin{eqnarray}
		h_{j,j+1}=\frac{1+\gamma^2}{2}(\sigma_j^+\sigma_{j+1}^-+\sigma_j^-\sigma_{j+1}^+),
	\end{eqnarray}
	and the jump operator
	\begin{eqnarray}
		l_{j,j+1}=\sqrt{\gamma/2}\left[(1-\gamma)(1-n_j)n_{j+1}-(1+\gamma)n_j(1-n_{j+1})\right.\nonumber\\
		\left.+i(\gamma-1)\sigma_j^-\sigma_{j+1}^+-i(\gamma+1)\sigma_j^+\sigma_{j+1}^-\right],
	\end{eqnarray}
	where $n_j=(1-\sigma_j^z)/2$ and $\gamma>0$ is a dimensionless parameter~\cite{ADD_1}. The model conserves total magnetization $S^z=\sum_j\sigma_j^z$, which serves as our $U(1)$ charge.
	
	The B3 model is related to a solution of braid Yang-Baxter equation $\check{R}(u,v)$ via $\mathcal{L}_{j,j+1}=\partial_v\check{R}_{j,j+1}(v,u)|_{v\to u}$~\cite{YB_2}. Such structure brings a set of conserved quantities $\{\tilde{Q}_\alpha\}$ in doubled Hilbert space satisfying $[\tilde{Q}_\alpha,\tilde{Q}_\beta]=[\mathcal{L},\tilde{Q}_\alpha]=0$. However, these operators are not physical charges in general. For example, it is customary to identify $\tilde{Q}_2=\sum_j\mathcal{L}_{j,j+1}$. As $\tilde{Q}_2$ does not correspond to a strong symmetry~\cite{Wangzhong}, no physical charge emerges from $\tilde{Q}_2$. Therefore, the only physical transport we study is that of $S^z$. The corresponding single-chain operators read	\begin{eqnarray}
		\frac{2A_{j,j+1}}{1+\gamma^2}=-\frac{\gamma}{2}(1-\sigma_j^z\sigma_{j+1}^z)+i(\gamma-1)\sigma_j^-\sigma_{j+1}^+\nonumber\\
		-i(\gamma+1)\sigma_j^+\sigma_{j+1}^-,
	\end{eqnarray}
	with $B_{j,j+1}=A^*_{j,j+1}$. We emphasize the asymmetric structure: particles hop right with amplitude $\propto(\gamma+1)$ and left with amplitude $\propto(\gamma-1)$. This asymmetry, inherited from the chiral Hamiltonian density, will prove essential. When $\gamma=1$, the B3 model reduce to the asymmetric simple exclusion process (ASEP) thus naturally having KPZ dynamics~\cite{Jin_1}, different from diffusion given by SSBA~\cite{SUP}. When $\gamma\neq1$, the coherent part $J_j^0$~\cite{YB_1} makes the dynamics complicated and recent numerical study suggests that it is diffusive~\cite{YB_BA}. We will show that KPZ dynamics still exists even when $\gamma\neq1$.

As noted previously, B3 model becomes two interacting spin-1/2 chains under Choi isomorphism. We find that for an initial state of  $\rho_0=2n_{j'}\ket{\Psi}\bra{\Psi}n_{j'}$, where
\begin{eqnarray}
		\ket{\Psi}= \otimes_{j=1}^N\left(\ket{0_j}+i^j\ket{1_j}\right)/\sqrt{2}
	\end{eqnarray}
satisfies
\begin{equation}
l^*_{j,j+1}\ket{\Psi^*}=0,
\end{equation}
the two chains become decoupled when the initial state is $\sigma_{j',l}^z\ket{\Psi}_l\Ket{\Psi^*}_r$.
The transport dynamics can be characterized by the corresponding evolution of modified charge and current densities
	\begin{eqnarray*}
		&&G(x=j-j',t)=\mathrm{Tr}\left(\rho_0\sigma_j^z(t)\right)-\mathrm{Tr}\left(\ket{\Psi}\bra{\Psi}\sigma_j^z(t)\right),\\
		&&\mathcal{F}(x,t)=\mathrm{Tr}\left(\rho_0J_j(t)\right)-\mathrm{Tr}\left(\ket{\Psi}\bra{\Psi}J_j(t)\right).
	\end{eqnarray*}
For a more detailed study, we decompose the charge correction function as
\begin{equation}
    G=-\frac{1}{2}\mathcal{G}_u-\frac{1}{2}\mathcal{G}_w+\frac{1}{2}\mathcal{G}_v,
\end{equation}
where
\begin{eqnarray}
&&\mathcal{G}_u=(\textbf{1}|\sigma_{j,l}^ze^{t\mathcal{L}}\sigma_{j',l}^z\ket{\Psi}_l\Ket{\Psi^*}_r,\\
&&\mathcal{G}_w=(\textbf{1}|\sigma_{j,l}^ze^{t\mathcal{L}}\sigma_{j',r}^z\ket{\Psi}_l\Ket{\Psi^*}_r,\\
&&\mathcal{G}_v=\mathrm{Tr}\left(\sigma_j^z(t)\sigma_{j'}^z\ket{\Psi}\bra{\Psi}\sigma_{j'}^z-\sigma_j^z(t)\ket{\Psi}\bra{\Psi}\right).	
\end{eqnarray}
Here, the terms $\mathcal{G}_u$ and $\mathcal{G}_w=\mathcal{G}_u^*$ are the Green's function of $\sum_{j}A_{j,j+1}$ and $\sum_{j}A^*_{j,j+1}$, respectively, probing the decoupled single-chain dynamics, while $\mathcal{G}_v$ involves genuine two-chain correlations. As we have pointed out earlier, SSBA fails to describe $\mathcal{G}_u$ and $\mathcal{G}_w$.

\smallskip{}
\noindent \textit{\textbf{Numerical results}}.---
Using time-evolving block decimation (TEBD) method~\cite{TEBD_1,TEBD_2} we numerically calculated the charge transport with a chain length of $N=256$, bond dimension $\chi=64$, and $\gamma=1.2$. Figure~\ref{fig_B3_lind}(a) shows the decay of charge significantly distinct from diffusive transport with $G(0,t)\sim t^{-1/2}$. Instead, the results fitts by a faster decay with $G(0,t)\sim t^{-2/3}$, confirming the KPZ scaling behavior~\cite{KPZ_scaling_function,website} with a dynamical exponent $z=3/2$, which emerges when $t\gtrsim5$. The distinction is further revealed by the scaling function, as shown in Figs.~\ref{fig_B3_lind}(b) and (c). The numerical results match universal KPZ scaling functions $f_{\mathrm{KPZ}}$ through the following ansatz
\begin{eqnarray}
G(x,t)&\simeq -\frac{b}{t^{2/3}}f_{\mathrm{KPZ}}(\frac{ax}{t^{2/3}}),\label{KPZ_a}\\
\mathcal{F}(x,t)&\simeq-\frac{2bx}{3at^{5/3}}f_{\mathrm{KPZ}}(\frac{ax}{t^{2/3}}),\label{KPZ_charge}
\end{eqnarray}
with fitted parameters $a=b\simeq0.426$ when $\gamma=1.2$.
	
Figure~\ref{fig_2} further studies the components $\mathcal{G}_{u,w,v}$ to reveal the origin of the KPZ dynamics. We find that $\mathcal{G}_{u,w}$ follows the KPZ charge scaling of Eq.~(\ref{KPZ_a}). In contrast, Fig.~\ref{fig_2}(b) shows that the two-chain component obeys the KPZ current scaling of Eq.~(\ref{KPZ_charge}) $\mathcal{G}_v(x,t)\simeq-\frac{2cx}{3dt^{5/3}}f_{\mathrm{KPZ}}(dx/t^{2/3})$ with $c\simeq0.08$ and $d\simeq0.4$, which decays as $t^{-5/3}$ and is faster than the $t^{-2/3}$ of $\mathcal{G}_{u,w}$. Consequently, $\mathcal{G}_v$ becomes negligible at long times. Therefore, the KPZ dynamics originate from $\mathcal{G}_{u,w}$, since $t^{2/3}G(x,t)\simeq -bf_{\mathrm{KPZ}}(a\xi)-\frac{c\xi}{3dt^{1/3}}f_{\mathrm{KPZ}}(d\xi)\to-bf_{\mathrm{KPZ}}(a\xi)$ when $t\to\infty$, with $\xi=x/t^{2/3}$. 
The results for $\gamma=0.5$ show the same structure, implying KPZ dynamics emerges regardless of $\gamma$~\cite{SUP}.

As mentioned above, the SSBA cannot reproduce these KPZ dynamics and instead predicts diffusive behavior. We further study the spectral properties of the Liouvillian to check whether it deviates from the diffusive signature. Surprisingly, this is not the case. The spectral gap in the $(Q_l,Q_r)=(m,m)$ subspace, which is the second biggest real part among the eigenvalues of $-\mathcal{L}^\dagger$~\cite{GAP}, coincides with that reported in a recent numerical study~\cite{YB_BA}. A direct calculation gives~\cite{SUP}
\begin{eqnarray}
	E_{k,q=m/N}\simeq2\gamma q(1-q)\left(-4i\gamma\left(1-2q\right)k+(1+\gamma^2) k^2\right)\label{Energy_B3_lind}.
\end{eqnarray}
Since the smallest allowed momentum is $k_\mathrm{min}=2\pi/N$, the spectral gap reads
\begin{eqnarray}
	\Delta_q=\mathrm{Re}(E_{2\pi/N,q})=8\pi^2q(1-q) \gamma(1+\gamma^2)N^{-2},\label{gap_NEW}
\end{eqnarray}
which scaling $\Delta\propto\gamma(1+\gamma^2)/N^2$ also agrees with the numerical results ~\cite{YB_BA,GAP_ADD}. The scaling of the gap corresponds to a diffusive dynamic component $z=2$, which is consistent with the diffusion behavior prediction by the SSBA. These results indicate that the SSBA only captures the relaxation spectrum of the full Liouvillian while failed in the characterization of the relevant charge transport governed by the single-chain dynamics, which is governed by a distinct gap as derived below.

\smallskip{}
\noindent \textit{\textbf{Emergent asymmetric XXZ structure}}.---
We now demonstrate analytically the origin of the KPZ dynamics by showing that the single-chain dynamics reduce to the asymmetric XXZ model~\cite{AS_XXZ_1,AS_XXZ_2,KPZ_sta} whose spectral gap $\Delta\propto N^{-3/2}$. Introducing $D=\otimes_{j=1}^N\mathrm{diag}(1,(-i)^j)$ and $|\phi_t)=D_l\otimes D_r^*e^{t\mathcal{L}}\sigma_{j',l}^z\ket{\Psi}_l\ket{\Psi^*}_r$, we obtain~\cite{SUP}
	\begin{eqnarray}
		\frac{d|\phi_t)}{dt}=\sum_{j,j+1}DA_{j,j+1}D^\dagger\otimes\textbf{1}|\phi_t)=-{H}_{\mathrm{XXZ}}\otimes\textbf{1}|\phi_t),
	\end{eqnarray}
	where
	\begin{eqnarray}
		\frac{2{H}_{\mathrm{XXZ}}}{1+\gamma^2}=\sum_{j=1}^{N} [-(\gamma+1)\sigma_j^+\sigma_{j+1}^--(\gamma-1)\sigma_j^-\sigma_{j+1}^+\nonumber\\
		+\frac{\gamma}{2}\left(1-\sigma_j^z\sigma_{j+1}^z\right)].
	\end{eqnarray}
	$\mathcal{G}_u$ thus becomes the Green's function of the statistical process described by ${H}_{\mathrm{XXZ}}$:
	\begin{eqnarray}
		\mathcal{G}_u=(\textbf{1}|\sigma_{j,l}^z|\phi_t)=\bra{\Phi}\sigma_j^ze^{-tH_{\mathrm{XXZ}}}\sigma_{j'}^z\ket{\Phi},\label{G_u}
	\end{eqnarray}
	where $\ket{\Phi}=D\ket{\Psi}=\otimes_{j=1}^N\frac{1}{\sqrt{2}}\left(\ket{0}_j+\ket{1}_j\right)$. When $\gamma>1$ $H_{\mathrm{XXZ}}$ describes ASEP. Therefore, we focus on the case $\gamma<1$, where the hopping rate $\gamma-1$ becomes negative. Although the KPZ equation cannot be derived directly from Eq.~(\ref{G_u}), its dynamics can be captured roughly at the mean field level, following an approach similar to Ref.~\cite{KPZ_sta}. Specifically, we find that~\cite{SUP}
	\begin{eqnarray}
		\frac{2}{1+\gamma^2}\partial_t\mathcal{G}_u(x,t)=\bra{\Phi}\left[-\gamma(2\sigma_j^z-\sigma_{j-1}^z-\sigma_{j+1}^z)\right.\nonumber\\
		\left.+\sigma_j^z(\sigma_{j+1}^z-\sigma_{j-1}^z)\right]e^{-tH_{\mathrm{XXZ}}}\sigma_i^z\ket{\Phi}.\label{KPZ1}
	\end{eqnarray}
	Expanding $q_{y=j\pm1}(t)=e^{tH_{\mathrm{XXZ}}}\sigma_{j\pm1}^ze^{-tH_{\mathrm{XXZ}}}$ to ${q}_{y=j}^z(t)\pm \delta\partial_y q_y(t)+\frac{1}{2}\delta^2\partial_y^2q_y(t)$, the equation above reduces to
\begin{eqnarray}
		\left\langle\frac{2}{1+\gamma^2}\partial_tq_y(t)-\delta^2\gamma\partial_y^2q_y(t)-2\delta q_y(t)\partial_yq_y(t)\right\rangle=0,
	\end{eqnarray}
with $\langle \cdot\rangle=\bra{\Phi}\cdot\sigma_i^z\ket{\Phi}$. As this equation is very similar to noise averaged Burgers equation, KPZ scaling emerges naturally. Notably, even though the hopping rate is negative, the diffusion constant $\delta^2\gamma$ remains positive, implying that the Burgers equation still serves as the effective theory.

To analytically prove that dynamical exponent $z=3/2$, we calculate the spectral gap of $H_{\mathrm{XXZ}}$ via the Bethe ansatz. For $\gamma>1$, the Bethe equations were solved in Ref.~\cite{AS_XXZ_1}, and the solutions remains valid for $\gamma<1$. The gap is
\begin{eqnarray}
		\Delta_{q=m/N} \simeq(1+\gamma^2) Y\sqrt{2q(1-q)}\left(\frac{\pi}{N}\right)^{3/2}+O(N^{-2}),\label{XXZ_gap}
\end{eqnarray}
where $Y$ is a constant irrelevant to $\gamma,\,N$ and $q$. The scaling $\Delta\propto N^{-3/2}$ confirms $z=3/2$~\cite{GAP}, in contrast to the $N^{-2}$ gap of the full Liouvillian discussed above [Eq.~\ref{gap_NEW}].

The KPZ behavior of current in Fig.~\ref{fig_B3_lind} (c) can be understand in the same way as that of charge discussed above. First, we introduce the single-chain current $\mathcal{F}_u$ and inter-chain current $\mathcal{F}_v$:
	\begin{eqnarray}
		&&\mathcal{F}_u(x,t)=(\textbf{1}|J_{j,l}^ze^{t\mathcal{L}}\sigma_{j',l}^z\ket{\Psi}_l\Ket{\Psi^*}_r,\label{J_u}\\
		&&\mathcal{F}_v(x,t)=\mathrm{Tr}\left(J_j(t)\sigma_{j'}^z\ket{\Psi}\bra{\Psi}\sigma_{j'}^z-J_j(t)\ket{\Psi}\bra{\Psi}\right),\nonumber
	\end{eqnarray}
so that $\mathcal{F}=-\mathcal{F}_u+\frac{1}{2}\mathcal{F}_v$. The currents of Burgers equation emerges~\cite{SUP}:
	\begin{eqnarray}
&&\mathcal{F}_u(x,t)=\bra{\Phi}\tilde{J}_je^{-tH_{\mathrm{XXZ}}}\sigma_{j'}^z\ket{\Phi},\\
		&&\tilde{J}_j=-\frac{1+\gamma^2}{2}\left(\gamma(\sigma_{j+1}^z-\sigma_j^z)+\sigma_{j}^z\sigma_{j+1}^z\right)\label{tilde_J}.
	\end{eqnarray}
Since $\tilde{J}_j\sim-\frac{1+\gamma^2}{2}\left(a\gamma\partial_yq_y+q_y^2\right)$, the single-chain current follows the KPZ current scaling $\mathcal{F}_u\simeq -\frac{2bx}{3at^{5/3}}f_{\mathrm{KPZ}}(ax/t^{2/3})$. By contrast, Fig.~\ref{fig_2}(c) shows that $\mathcal{F}_v$ does not scales as the Eq.~(\ref{KPZ_charge}). Although $\mathcal{F}_v$ contributes to $\mathcal{F}$, its magnitude is negligible compared to $\mathcal{F}_{u}$. Consequently, the scaling of $\mathcal{F}(x,t)$ closely aligns with Eq.~(\ref{KPZ_charge}), confirming that the decoupled asymmetric XXZ chain contribute to the KPZ dynamics for both charge and current.
	
\smallskip{}
\noindent \textit{\textbf{Discussion}}.---
In summary, we have identified a hidden asymmetric XXZ structure within the B3 model, which gives rise to KPZ dynamics for suitable initial states that beyond the reach of the SSBA. This demonstrates that a phenomenological understanding based on SSBA is insufficient to capture the emergent hydrodynamics of open systems. Several questions remain open: First, it is unclear how to analytically explain the behavior of $\mathcal{G}_v$ and $\mathcal{F}_v$. Second, it remains unclear whether KPZ dynamics emerges for other classes of initial states. Addressing these questions may benefit from the nested Bethe ansatz method applied to the B3 model~\cite{YB_3,YB_BA}, and could be probed experimentally with the advances in experimental quantum simulation platforms, e.g., the neutral atom array~\cite{Weimer2010,Browaeys2020,Ma2026}.

More broadly, our work reveals a pathway for connecting classical statistics with open quantum systems: if $\textbf{1}\otimes l_{j,j+1}^*|\rho)=\textbf{1} \otimes B_{j,j+1}|\rho)=0$ hold and $\sum_j A_{j,j+1}$ is proportional to the generator of some statistical model, the correlation function $\mathrm{Tr}[\mathcal{O}'(t)\mathcal{O}(0)\rho]$ coincides with the Green's function of that classical process. This correspondence allows understanding dynamics of many-body systems through the developed tools and results achieved in classical statistics. The B3 model provides a concrete example: it reduces to the ASEP in this manner when $\gamma\geq1$. We anticipate that the generalization of more interesting classical statistical models to open quantum systems can be stimulated by this work. Finally, we note that the effective  hopping rate becomes negative for $\gamma<1$. Such negative-rate Markov processes have been introduced only very recently~\cite{NMP_1, NMP_2, NMP_3}, and our construction supplies a new physical origin. Because negativity offers possible novel scaling behavior, it would be interesting to find them in open quantum systems.

\begin{acknowledgments}
We thank Qing-Xuan Jie and Bo-Tao Gao for inspiring discussions. This work was funded by the National Key R\&D Program (Grant No.~2021YFA1402004), and the National Natural Science Foundation of China (Grants No.~92465201 and 92265210).  This work was also supported by the Fundamental Research Funds for the Central Universities and USTC Research Funds of the Double First-Class Initiative. The numerical calculations in this paper were performed on the supercomputing system in the Supercomputing Center of USTC.
\end{acknowledgments}


\begin{thebibliography}{51}
\bibitem [{ADD()}]{ADD_1}%
  \BibitemOpen
  \href@noop {} {\ }\bibinfo {note} {Lindblad master equation is invariant
  under transformation $(h,l)\to(h+i(xl^\dagger-x^*l),l+2x)$ for any complex
  number $x$. Here we just choose $x=-\frac{\gamma}{2}\sqrt{\frac{\gamma}{2}}$
  comparing with the original model.}\BibitemShut {Stop}%
\bibitem [{GAP()}]{GAP_ADD}%
  \BibitemOpen
  \href@noop {} {}\bibinfo {note} {The extra factor $(1+\gamma^2)$ is from the
  fact that our B3 model transform to the B3 model in \cite{YB_BA} via
  $(h,l)\to((1+\gamma^2)^{-1}h,(1+\gamma^2)^{-1/2}l)$.}\BibitemShut {Stop}%
\end{thebibliography}
\end{document}